\documentclass[prl,showpacs,twocolumn,amsmath,amssymb,groupadress,superscriptaddress]{revtex4}
\usepackage{graphicx}% Include figure files
\usepackage{natbib}
\newcommand{\rmno}{RMn$_2$O$_5$ }

\newcommand{\Y}{YMn$_2$O$_5$ }

\newcommand{\Tbun}{TbMnO$_3$ }
\newcommand {\Mt} {Mn$^{3+}$ }
\newcommand {\Mf} {Mn$^{4+}$ }

\begin{document}

\title{The incommensurate magnetic structure of YMn$_2$O$_5$: a stringent test of the multiferroic mechanism}

\author{P.G. Radaelli}
\affiliation{ISIS facility, Rutherford Appleton Laboratory-STFC,
Chilton, Didcot, Oxfordshire, OX11 0QX, United Kingdom. }
\affiliation{Dept. of Physics and Astronomy, University College
London, Gower Street, London WC1E 6BT, United Kingdom}
\author{C. Vecchini}
\affiliation{ISIS facility, Rutherford Appleton Laboratory-STFC,
Chilton, Didcot, Oxfordshire, OX11 0QX, United Kingdom. }
\affiliation{Institute of Electronic Structure and Laser, Foundation for Research and Technology - Hellas, Vassilika Vouton, 711 10 Heraklion, Crete, Greece. }
\author{L.C. Chapon}
\affiliation{ISIS facility, Rutherford Appleton Laboratory-STFC,
Chilton, Didcot, Oxfordshire, OX11 0QX, United Kingdom. }
\author{P.J. Brown}
\affiliation{Institut Laue-Langevin, 6, rue Jules Horowitz, BP 156 - 38042 Grenoble Cedex 9 - France.}
\author{S. Park}
\affiliation{Department of Physics and Astronomy, Rutgers
University, Piscataway, New Jersey 08854, USA}
\author{S-W. Cheong}
\affiliation{Department of Physics and Astronomy, Rutgers
University, Piscataway, New Jersey 08854, USA}
\date{\today}% It is always \today, today,

\begin{abstract}
We have determined the magnetic structure of the low-temperature incommensurate phase of multiferroic YMn$_2$O$_5$ using single-crystal neutron diffraction.  By employing corepresentation analysis, we have ensured full compliance with both symmetry and physical constraints, so that the electrical polarization must lie along the $b$ axis, as observed.  The evolution of the spin components and propagation through the commensurate-incommensurate phase boundary points unambiguously at the exchange-striction mechanism as the primary driving force for ferroelectricity.
\end{abstract}


\maketitle

%\begin{figure}[!h]
%\includegraphics[scale=0.40, angle=-90]{PRL_YMn2O5_Radaelli_Figure1.eps}
%\caption{(color online)}
%\end{figure}

\begin{figure}[h!]
\includegraphics[scale=0.37, angle=-90]{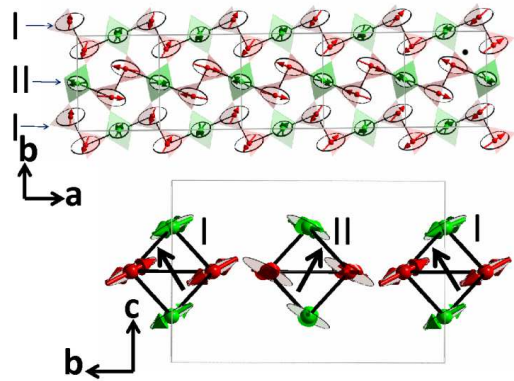}
\caption{\label{Fig: LT-ICM} (color online) Low-temperature incommensurate magnetic structure of \Y. \textbf{Top}:  projection of the magnetic structure onto the $ab$-plane. \Mt and \Mf are shown with red and green color, respectively. The ellipsoidal contours, more elongated for \Mt, show the envelop of the magnetic modulations. Adjacent AFM zig-zag chains, labelled I and II, are shown by thick black lines. \textbf{Bottom}:  Projection of the magnetic structure onto the $bc$-plane. The black arrows represent the normal direction to the ellipoidal modulations in chain I and II.}
\end{figure}

The family of compounds with general formula \rmno ($R$=Y, Rare earth, Bi and La) \cite{ISI:000221644600033} occupies a special place among the so-called "novel" multiferroic materials, in which ferroelectricity is directly induced by the onset of inversion-breaking magnetic ordering.  In fact, whereas for most other multiferroics, such as \Tbun \cite{ISI:000186370800038, ISI:000231310900063} Ni$_3$V$_2$O$_8$ \cite{ISI:000225661100073} and many other materials, the so-called spin-orbit (SO) mechanism \cite{ISI:000235394100083, ISI:000250620300074} has clearly been identified as the primary driving force for the electrical polarization, the origin of ferroelectricity in \rmno remains controversial.  Three aspects of the \rmno phenomenology suggest that the SO mechanism may not play a major role :  the direction of the electrical polarization cannot be rotated by an applied magnetic field \cite{ISI:000231564400088}; the main ferroelectric phase is magnetically \emph{commensurate} while the low-temperature incommensurate phase (LT-ICP) is only weakly ferroelectric,  and, most importantly, in the ferroelectric commensurate phase (CP), moments in the \textit{ab}-plane are almost collinear (non-collinearity is a strict requirement of the SO mechanism). An alternative explanation, based on the simple exchange-striction (ES) mechanism in the context of an acentric quasi-collinear structure, has been shown to account qualitatively for the ferroelectric behavior of the different magnetic phases \cite{ISI:000224662700076, Betouras_07}, and recent electronic structure calculations not including the SO interaction \cite{bodenthin:027201} have reproduced the observed value of the electrical polarization $P$ for certain values of the parameters.  However, up to now the SO mechanism could not be completely ruled out, since accurate neutron diffraction determinations of the CP magnetic structure \cite{vecchini, ISI:000248244500035} have identified a small cycloidal component in the ferroelectric phases.\\
\indent One striking experimental fact about \rmno is the sudden, dramatic reduction in the value of $P$ at the CP to LT-ICP transition. Understanding how the different components of the magnetic structure change could provide key insight into the multiferroic mechanism. More specifically, we have shown \cite{Radaelli_IOP_08} that, starting from the experimental magnetic structure, one can construct time-reversal-even polar vectors \emph{specific} to each mechanism, which are directly proportional to the SO and ES contributions to $P$. Most likely, the coupling constants do not change across the transition; therefore, the magnitude of the electrically active polar vector \emph{must}, like $P$, be greatly reduced across the transition. For this approach to be valid, the magnetic structures must be known with great accuracy. In this respect, \Y is an ideal model system since there is no rare-earth magnetic ordering involved.  At present, whereas the \Y CP magnetic structure is well established, and has been validated independently by two groups \cite{vecchini, ISI:000248244500035}, the LT-ICP structure is not known with confidence.  The two structures presented to date from neutron powder diffraction \cite{ISI:000235905700072}, and more recently, from single crystal data \cite{kim} are significantly different, and, crucially, neither of them exploits the full set of symmetry and physical constraints, requiring that the residual electrical polarization in the LT-ICP remains parallel to the $b$ axis.\\
\indent Here, we present a single crystal neutron diffraction determination of the \Y incommensurate magnetic structure that is fully compliant with symmetry and physical constraints.  In particular, we have employed corepresentation analysis to impose the $\cdot 2 \cdot$ point-group symmetry, which is lower than the CP symmetry ($m2m$) but still guarantees that all the contributions to $P$ are directed along the $b$ axis, as observed experimentally.  The fit to the data, collected on two independent magnetic domains, is much better that for previous structures \cite{ISI:000235905700072, kim}, and is essentially equivalent to our best unconstrained refinement.  The spins responsible for the ES polarization are now found to be nearly perpendicular and are modulated with opposite phases, leading to a reduction of the ES polar vector by a factor of ~9 with respect to the CP.  On the contrary, the magnitudes of the SO polar vectors \emph{increase} significantly in the LT-ICP, while a new polar vector of SO origin and of smaller magnitude develops due to the formation of long-period cycloids along the $a$ axis.  Barring an accidental cancelation of SO terms, which is extremely unlikely to occur for all \rmno compounds and all temperatures, this proves conclusively that ferroelectricity in the \rmno commensurate phases is predominantly of exchange-strictive origin.  However, it is possible that the SO mechanism may provide a contribution to the residual polarization in the low-temperature incommensurate phase.\\
\indent The \Y single crystal for this study is the same as for reference \cite{vecchini}, and the growth protocol is described therein.  The experimental setup was also the same but the data were collected at 2K.  The main difference between the experiments on the CP and LT-ICP phases is at the data reduction stage:  in the CP with propagation vector $k=2\pi(\frac{1}{2}, 0, \frac{1}{4})$, there are two inversion-related domain that scatter onto the same points in reciprocal space, yielding single Bragg peaks on the position-sensitive detector (PSD). On the other hand, the LT-ICP has 4 distinct domains related by inversion and/or rotation around the $c$ axis. The rotation-related domains have propagation vectors $k_1=2\pi(0.48, 0, 0.28)$ and $k_2= 2\pi(-0.48, 0, 0.28)$.  Scattering from these domains occurs at close positions in reciprocal space, yielding split peaks on the PSD. An \textit{ad hoc} piece of software was developed to de-convolute the peaks from the two domains and exclude the reflections that were completely overlapped. Because of the different integration methods employed for the CP and LT-ICP, absolute scaling of the magnetic moments was verified using the neutron powder diffraction data reported in \cite{ISI:000235905700072}.  The FullProf program \cite{fullprof} was used for the refinements of the magnetic structure.  In the final analysis, reflections from the two domains were combined in a single refinement, with the spin components of the two domains constrained by symmetry (see below).\\
\indent For the \textit{corep} analysis, we use the Kovalev conventions as in \cite{ISI:000249155100099}.
The propagation vector is $k=2\pi(0.48, 0, 0.28)$, ($k_{3}$ in the Kovalev notation).
The little \textit{irrep} group contains two operators --- the identity $h_1$ and the $a$-glide $h_{27}$ perpendicular to $b$.  According to Kovalev (see Table \ref{tab:Pbam_irreps}), there are only two \textit{irreps}, both 1-dimensional, each generating a \textit{corep}. Both sites, Mn$^{3+}$ and Mn$^{4+}$, split into two orbits in the little group, which are then recombined by the \textit{coreps}.  The transformation matrix $\beta$ is the identity. The axial-vector modes were symmetrized with respect to the center of symmetry located at position $0,0.5, 0.5$.  The two-fold screw $h_3$ is located at $0.5, y, 0.5$. In this respect we depart from Kovalev's conventions.
\begin{table}[h!]
\caption{\label{tab:Pbam_irreps}  Small \textit{irreps} ($\Delta$)
and \textit{coreps} ($D$) of space group $Pbam$ for propagation
vector $k_3=(\mu,0,\mu)$. The symmetry operators are in the Kovalev
notation.  $\epsilon=e^{-2\pi i\,\frac{k_x}{2}}=0.0628-i\,0.998)$.}
\begin{tabular}{c|cccc}
& $h_1$& $h_{27}$ & $Kh_{25}$ & $Kh_3$\\
\hline
$\Delta_1/D_1$&1& $\epsilon$ & 1 & $\epsilon$ \\
$\Delta_2/D_2$&1& -$\epsilon$ & 1 & -$\epsilon$\\
\end{tabular}
\end{table}
Magnetic structures built from single-\textit{corep} modes are by construction invariant by inversion ($h_{25}$), and are also invariant by rotation around the $b$ axis ($h_3$) and reflection through a plane perpendicular to the $b$-axis ($h_{27}$), except for the phase factor $\epsilon$ or $-\epsilon$, which, in an incommensurate structure, is always equivalent to a translation and does not affect the point-group symmetry. Our aim is to construct a magnetic structure that allows the development of polarization along the $b$ axis--- we want to drop the $h_{25}$ and $h_{27}$ invariance but retain the $h_3$ rotation around $b$ (point group symmetry $\cdot 2 \cdot$).  Inspection of Table \ref{tab:Pbam_irreps} immediately suggest that this is accomplished by the combination $D_1+iD_2$.  In fact, inversion and rotation appear in the little \textit{corep} group as  antiunitary operators.  The imaginary unit in front of $D_2$ yields a sign change for these operators, ensuring the correct symmetry behavior.\\
\indent
As usual, each \textit{corep} mode for a single spin component has two parameters --- an amplitude and a phase, the latter describing the phase difference between inversion-related sites.  Therefore, the most generic magnetic structure invariant by $\cdot 2 \cdot$ symmetry is described by 24 parameters (2 parameters $\times$ 3 spin components $\times$ 2 \textit{coreps} $\times$ 2 sites).  However, this number can be further reduced by imposing \textit{physical} constraints.  Sites related by inversion in the paramagnetic phase have almost identical magnetic environments (except for the tiny ferroelectric displacements). Hence, we expect not only that the wave-amplitudes will be identical, but also that the spins follow the same propagation \textit{minus} a phase factor. This reduces the number of phases per site from 6 to 1 and the total number of parameters to 14.  One of these parameters represents the overall phase and can also be fixed (we have chosen to fix $ub^2_x=0$, see below), yielding 13 parameters in total, a drastic reduction from 48 for an unconstrained refinement \cite{kim}.\\
The $j$ component of the magnetic moment on site $l$ and unit cell $n$ (at a position $R_n$ from the origin) is written as
\begin{equation} \label{Eq: Fourier_expansion}
m_j^l(n)=S_{j}^l\,e^{ik \cdot R_n} + c.c.
\end{equation}
The Fourier coefficients have been parameterized as, for example for component $x$ of site $b$:
\begin{equation} \label{Eq: Example_coeff}
S_{x}^b=\frac{1}{2}(ub^1_x+iub^2_x)\,e^{i \Phi}
\end{equation}
where $\Phi$ is a global phase common to all components.  The refined parameters for Domain 1 and the appropriate transformation to obtain those of Domain 2 are listed in Table \ref{tab:Domain1_mode}.
\begin{table}[h!]
\caption{\label{tab:Domain1_mode} \Y LT-ICP magnetic structure at T = 2 K. Parameterized Fourier coefficients for Domain 1, $k_1=2\pi(0.48, 0, 0.28)$. The refined values are:  $ub^1_x=-2.45(3)$, $ub^1_y=0.30(6)$, $ub^1_z=-0.27(10)$, $ub^2_x=0$, $ub^2_y=1.72(1)$, $ub^2_z=1.03(3)$, $\psi_b=-0.36(1)$, $ua^1_x=3.50(3)$, $ua^1_y=0.46(8)$, $ua^1_z=0.26(7)$, $ua^2_x=-0.53(8)$, $ua^2_y=-2.00(1)$, $ua^2_z=1.20(4)$, $\psi_a=-1.650(7)$.  $R_F=6.75/6.67$ \%\ and $R_F^2=10.7/9.83$ \%\ for Domain 1/Domain 2, respectively.  The transformations to obtain the Domain 2 parameters are: $k_2=2\pi(-0.48, 0, 0.28)$   $u^1_x \rightarrow -u^1_x$, $u^1_y \rightarrow -u^1_y$,$u^1_z \rightarrow u^1_z$,  $u^2_x \rightarrow -u^2_x$, $u^2_y \rightarrow -u^2_y$, $u^2_z \rightarrow u^2_z$ for both $a$ and $b$ sites, and $\psi_b \rightarrow \psi_b$, $\psi_a \rightarrow -\psi_a$. Labeling of atoms is as in \cite{vecchini}.  Supplementary Table I in the EPAPS contains a complete list of parameters for both domains, using the FullProf conventions.}
\begin{tabular}{|c|c|c|c|c|c|c|c|}
\hline
Site&\multicolumn{6}{c|}{Fourier coefficients} &  Global phase $\Phi$\\
\hline
b1&$+ub^1_x$ & $+ub^1_y $ & $+ub^1_z$ & $+ub^2_x$   & $+ub^2_y$ & $+ub^2_z$ & $\psi_b$\\
b2&$+ub^1_x$ & $+ub^1_y $ & $+ub^1_z$ & $+ub^2_x$   & $+ub^2_y$ & $+ub^2_z$ & $ -\psi_b$\\
b3&$-ub^1_x$ & $+ub^1_y$ &  $-ub^1_z$ & $+ub^2_x$   & $-ub^2_y$ & $+ub^2_z$ & $q_x/2+\psi_b$\\
b4&$-ub^1_x$ & $+ub^1_y$ &  $-ub^1_z$ & $+ub^2_x$   & $-ub^2_y$ & $+ub^2_z$ & $q_x/2-\psi_b$\\
a1&$+ua^1_x$ & $+ua^1_y $ & $+ua^1_z$ & $+ua^2_x$   & $+ua^2_y$ & $+ua^2_z$ & $\psi_a$\\
a2&$-ua^1_x$ & $+ua^1_y$ &  $-ua^1_z$ & $+ua^2_x$   & $-ua^2_y$ & $+ua^2_z$ & $q_x/2-\psi_a$\\
a3&$-ua^1_x$ & $+ua^1_y$ &  $-ua^1_z$ & $+ua^2_x$   & $-ua^2_y$ & $+ua^2_z$ & $q_x/2+\psi_a$\\
a4&$+ua^1_x$ & $+ua^1_y $ & $+ua^1_z$ & $+ua^2_x$   & $+ua^2_y$ & $+ua^2_z$ & $q_x-\psi_a$\\
\hline
\end{tabular}
\end{table}
The fit is far superior to that obtained using the model by Kim \textit{et al.} \cite{kim} ($R_F \simeq 6.7$ \%\ vs $15.5$ \%\ for Kim), and is better than any unconstrained refinement we could obtain based on a simulated annealing structure.
The magnetic structure of the LT-ICM is depicted in Fig \ref{Fig: LT-ICM}.  The top panel shows the $ab$-plane projection of the \emph{envelope} of the magnetic structure, i.e., several spin orientations describing a complete period of propagation are shown on the same site.  The bottom panel shows a perspective view of the magnetic structure illustrating the phase relations between spins on different sites.  The LT-ICM adopts the most generic type of single-$k$ structure, in which spins rotate in an arbitrary plane with an elliptical envelope --- in other words all spin components are involved in the cycloids.  These envelopes are strictly related by symmetry on the different \Mt and \Mf sites, and differ only by a phase factor.  \Mt and \Mf envelopes are almost co-planar, but the \Mt ellipse is much more eccentric (the semiaxes are 3.65 and 2.21 $\mu_B$ for \Mt and 2.48 and 2.00 $\mu_B$ for \Mf) , reflecting the different anisotropy of the two sites.  The \Mt long semiaxis coincides with the pyramid axis, consistent with the easy magnetic axis from susceptibility measurements and also with the results on the CP, whereas the short semiaxis points towards one of the oxygen atoms in the pyramid.  The most important aspect of this structure is the \emph{out of phase} relation between zig-zag chains running along the $a$ axis (bottom panel of Fig. \ref{Fig: LT-ICM}).  This has two consequences:  first, neighboring spins on different chains are essentially orthogonal;  second, even when the angle is not $90^{\circ}$ due to the eccentricity, the dot product across chains cancels out almost exactly due to the sign change in different part of the structure.  As the ES polarization is proportional to this dot product, we argue that this loss of phase coherence is the primary cause of the loss of ferroelectricity in the LT-ICP (see below). As in many other cycloidal magnets, the CP to LT-ICP transition can be well explained in terms of competition between anisotropy (which favors quasi-collinear arrangements) and entropy (which favors large-moment cycloids at low temperatures), but,  intriguingly, in \rmno the ferroelectric properties of the two phases are \emph{reversed} with respect to \Tbun \cite{ISI:000186370800038}.
In order to calculate the different contributions to the polarization, it is useful to extract modules ($m_i$) and phases ($\phi_i^u$) for the complex arrays $z_a$ and $z_b$, defined as $z_{ax}=ua^1_x+i\,ua^2_x$ etc.:
\begin{eqnarray}
\phi^u_a&=&\arg(z_a)\,;\,\phi^u_b=\arg(z_b)\nonumber\\
m_a&=&|z_a|\,;\,m_b=|z_b|
\end{eqnarray}
Using these equations, one can readily calculate all the macroscopic quantities of interest,
and in particular the polar vectors corresponding to the different mechanisms \cite{Radaelli_IOP_08}.  Since the
point-group symmetry is $\cdot 2 \cdot$, only the $y$ component of any polar vector will be
non-zero.  This can readily be verified by calculating the vector component explicitly, but it is not shown here.  By employing the transformations in Tab. \ref{tab:Domain1_mode}, we can also verify explicitly that all the polar vector change sign in Domain 2.  The formulas given here below are valid for the LT-ICM (no umklapp term \cite{Radaelli_IOP_08}), while for the CP we use eq. 3 and 4 in \cite{Radaelli_IOP_08}.
The ES polar vector is calculated as:
\begin{equation}
E_y= \sum_{j=1}^3m_{aj}\,m_{bj} \,[4\cos\psi_b\sin\psi_a\,\sin(\phi^u_{bj}-\phi^u_{aj})]
\end{equation}
The two spin-orbit components that were already present in the CP, originating from $bc$-plane cycloids and relating to SO interactions across the \Mt planes ($S^1_y$) and Y planes ($S^2_y$) are:
\begin{eqnarray}
S^1_y &=& m_{by}m_{bz}[-2\sin2\psi_b\sin(\phi^u_{by}-\phi^u_{bz})]\nonumber\\
S^2_y &=& m_{by}m_{bz}[2\sin(2\psi_b+q_z)\sin(\phi^u_{by}-\phi^u_{bz})]
\end{eqnarray}
In order to evaluate the spin-orbit contribution due to the in-plane projection of the cycloids (which was absent in the CP), we only consider the Mn$^{3+}$-Mn$^{4+}$ within the chains, since the other bonds have a small projection on the direction of propagation (spins on Mn$^{3+}$ atoms related by inversion are also almost antiparallel).
\begin{eqnarray}
&L_y=-4\sin(q_x/2-\psi_a)\cos\psi_b\cos\alpha \cdot\nonumber\\
&[m_{bx}m_{ay}\sin(\phi^u_{bx}+\phi^u_{ay})+m_{by}m_{ax}\sin(\phi^u_{by}+\phi^u_{ax})]
\end{eqnarray}
Table \ref{tab: FE} lists the polar-vector components for the CP \cite{vecchini} and LT-ICP, as determined from the previous formulas.  The relative signs of the different components are uniquely established, but the overall sign depends on the choice of domain.
\begin{table}[h!]
\caption{\label{tab: FE}  Exchange-striction ($E_y$) and spin-orbit ($S^1_y$, $S^2_y$ and $L_y$) polar vectors for the CP and LT-ICP phases of \Y.  All the values are in $\mu_B^2$.  Statistical errors are propagated from the magnetic structure refinements.}
\begin{tabular}{|c|c|c|c|c|}
\hline
Phase & $E_y$ & $S^1_y$ & $S^2_y$ & $L_y$  \\
\hline
CP & $22.6(5)$  &  $-0.16(4)$  & $-0.40(5)$  &   0\\
LT-ICP & $-2.5(3)$& $-1.01(8)$ & $-1.3(1)$ & $0.7(3)$\\
\hline
\end{tabular}
\end{table}
Tab. \ref{tab: FE} provides an essentially unambiguous test for the multiferroic mechanisms:  the drop of the ES component $E_y$ is by far the strongest candidate to account for the sudden loss of ferroelectricity at the CP/LT-ICP transition, as the SO components $S^1_y$ and $S^2_y$  significantly \emph{increase} in magnitude.  The only plausible SO-based alternative is an accidental cancelation between $S^1_y$, $S^2_y$ and $L_y$.  However, this would be an extraordinary coincidence, since the corresponding SO coupling constants may not even be of the same order of magnitude, as the exchange pathways are completely different.  Furthermore, this coincidence would be required for all the \rmno compounds and all temperatures, in spite of the significant perturbations introduced by the rare earth magnetism \cite{Beutier}.  The origin of the residual ferroelectricity in the LT-ICP is, however, a different matter.  The drop in $E_y$ ($\sim 89\%$) is somewhat larger that the drop in $P$ ($\sim 80\%$), making it plausible that the SO mechanisms may contribute significantly to the LT-ICP electrical polarization.\\
\indent In summary, we have determined the magnetic structure of the low-temperature incommensurate phase of \Y from single-crystal neutron diffraction data.  By employing corepresentation analysis, we have imposed strict symmetry and physical constraints, so that all component of the electrical polarization lie along the $b$-axis, as observed experimentally.  Quantitative analysis of the magnetic structure change across the commensurate-incommensurate transition points unambiguously to exchange-striction as the primary origin of ferroelectricity in this system.

%\bibliography{Radaelli_IC}

\end{document}